# Calibration of a superconducting transformer by measuring critical current of a NbTi Rutherford cable

Hui Yu, Jeremy Levitan and Jun Lu

National High Magnetic Field Laboratory, Florida State University, Tallahassee, Florida 32310

**Abstract**

Large high field superconducting magnets often requires high current superconducting cables. In order to develop these cables, a facility capable of providing high magnetic field with large sampling area as well as electrical current of tens of kA is essential. A superconducting transformer is an energy-efficient and low-cost way to provide large current to superconducting cables. Previously, we co-developed a superconducting transformer and successfully tested it to a maximum output current of 45 kA in zero magnetic field. In this work, this superconducting transformer is installed to the 12 T split solenoid magnet at the National High Magnetic Field Laboratory (NHMFL). We calibrated it by using this facility to measure critical current of a NbTi Rutherford cable as a function of magnetic field up to 10 T, and compare the results with those available in the literature. In addition, a strand extracted from the NbTi cable is tested for critical current. The critical current of the extracted strand is scaled and compared with critical current of the cable. The accuracy of the critical current measurement using this superconducting transformer is discussed in detail. This work concludes the commissioning of this superconducting transformer which combined with the 12 T split magnet will provide unique cable testing capability for future cable development for the NHMFL and its users.

1. Introduction

A superconducting cable that carries tens of kA of electrical current is usually used as conductor for large superconducting magnets such as accelerator magnets and fusion magnets. For the development of superconducting cables as well as the quality assurance testing of these cables, critical currents at their operating temperature and field need to be measured. Measurement of critical current of cables is usually performed in specialized testing facilities, of which a key component is a multi-kA DC current source. A superconducting transformer (SCT) is an efficient way to provide very high DC current at low temperatures for superconducting cable testing. SCT is a transformer with both primary and secondary windings made of superconductors. As a result of large primary to secondary turn-ratio, a small primary current can drive a very large secondary current as the output of the SCT. Compared with a conventional high current DC current source, SCT is preferred in superconducting cable testing facilities due to its lower cost, higher energy efficiency and lower liquid helium consumption. SCTs are installed as current sources at a number of facilities, such as University of Twente [1], the SULTAN facilities at the Centre de Recherches en Physique des Plasma (CRPP) [2], and the FRESCA at the CERN [3], [4].

The NHMFL has been developing an SCT in collaboration with the Lawrence-Berkley National Laboratory (LBNL). The SCT was designed and constructed initially at the LBNL [5]. In order to commision the SCT as an NHMFL user facility for superconducting cables research for future large superconducting magnets, we designed a new electronic control system for this SCT and tested in zero magnetic field up to 45 kA [6].

In this work, we installed this SCT inside the NHMFL 12 T split solenoid magnet cryostat. The proper functioning of the SCT is calibrated by measuring the critical current (Ic) of a LHC NbTi Rutherford cable as a function of magnetic field from 2 to 10 T. The LHC NbTi cable was chosen as the reference, because its superconducting property has been characterized and reported in the literature [xx]. The results are corrected for self-field which is calculated by the finite element method (FEM), and consistent with data in the literature. The accuracy of the measurement will be discussed in detail.

2. **Experimental Methods**

2.1 The superconducting transformer and its control

Fig. 1 (a) is a picture of the SCT which consists of a primary coil of 10,464 turns of NbTi wire and secondary coil of 6.5 turns of NbTi Rutherford cable [5]. Two Rogowski coils are installed as an integarted part of the SCT for output current measurement. In addition, two Hall sensors are attached to a demountable part near the output of the SCT. The Hall sensors are HZ-312C made by Asahi Kasei Corporation. Each of them are positioned at the center of each output lead, sensing the transverse field which is correlated with output current as shown in Fig. 1 (b). The Hall sensors were calibrated at room temperature and 4.2 K using a Quantum Design PPMS. The current measurement by the Hall sensors are verified at room temperature at about 200 A using a DC power supply. The current measured by Hall sensors were also calibrated at 4.2 K in zero field by the Rogowski coil which was in turn calibrated by a DC power supply up to 4 kA. These carefully calibrated Hall sonsors provide an additional mean of output current measurement.

A model 4G four-quadrant magnet power supply made by Cryomagnetics was used to control the primary current. This power supply also provides the detection and protection against a SCT or sample quench. For maximum stored mangetic energy of 45 kJ, no additional quench protection is needed as cofirmed by our previous tests [6]. The data acquisition system includes National Instruments SCXI-1000. The signals from the Rogowski coils, Hall sensors, and voltage taps on the sample are digitized by NI SCXI-1125 data input module and recorded by a LabVIEW program at a rate of 500 data points per second.

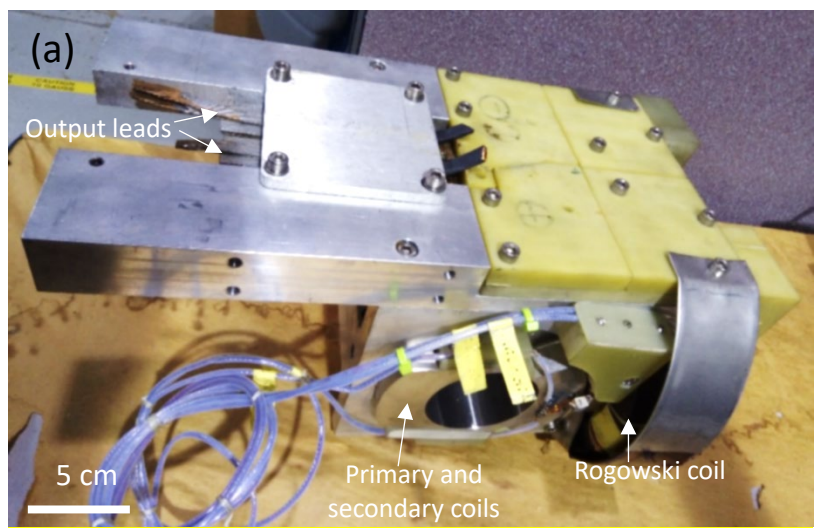

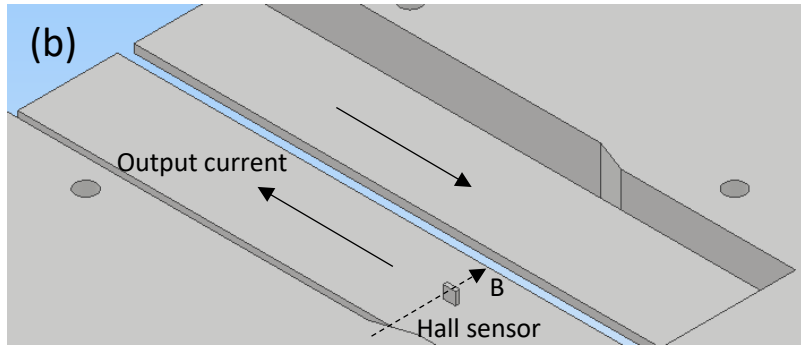

Fig. 1. (a) SCT used in this research, and (b) a schematic of the Hall sensor location and orientation.

2.2 The sample

The sample is an LHC type-01 NbTi Rutherford cable. The geometric dimensions of this cable are listed in Table I [7]. The strands in the cable are soldered together.

TABLE I  Dimensions of the LHC type-01 Rutherford cable [7]

| Dimensions | Value |
| --- | --- |
| Width | 15.0 mm |
| Thin edge | 1.72 mm |
| Thick edge | 2.06 mm |
| Number of strands | 28 |
| Strand diameter | 1.065 mm |
| Tranposition pitch | 110 mm |

2.3  Design of the test probe and preparation of the joints to SCT

The critical current of the sample may be measured in one of the two configurations, namely applied field parallel or perpendicular to the broad face of the cable. We decided to use the latter configuration, because it tends to give a more conservative $I_c$ value. A test probe was designed and constructed for this experiment as shown in Fig. 2.

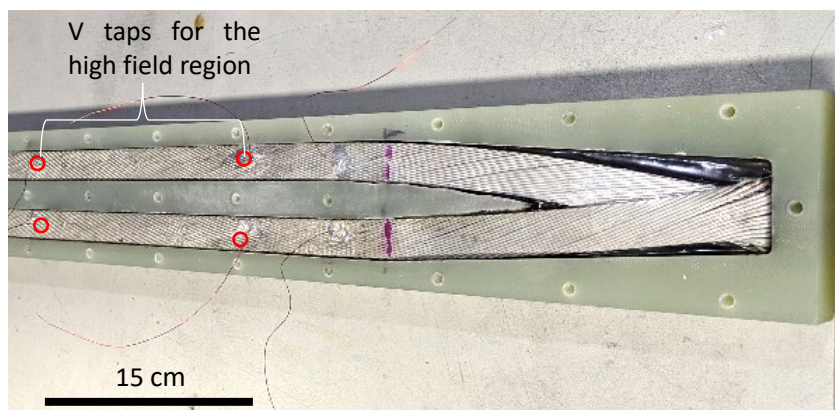

Fig. 2. The test probe. The Rutherford cable sample with its broad face glued down in a groove by Stycast 2850FT. The voltage tap distance is 15 cm, consistent with the bore size of the split magnet used for the experiment. The applied magnetic field is perpendicular to the broad face of the sample.

The probe is made of G-10 with machined grooves where the sample is glued by Stycast 2850FT. The sample has a V-shaped sharp bend at the return outside the high field region. Our experiment proved that this sharp bend in low field did not cause any degradation that prevents good $I_c$ measurements. When the probe cover (not shown) is glued by Stycast and bolted, the sample is completely fixed in the probe. This minimizes the possible movement of the sample due to electromagnetic force during a $I_c$ measurement, therefore minimizing the probability of a sample quench. The cross-section of the probe is 30 mm x 70 mm$^2$.

A soldering fixture and a soldering procedure have been developed to make solder lap join between the sample and the SCT output. The solder is eutectic $Sn_{63}Pb_{37}$. The joint region is about 5 cm long. Subsequently the assembly of the sample probe and the SCT was installed into the split magnet outside its cryostat as shown in Fig. 3.

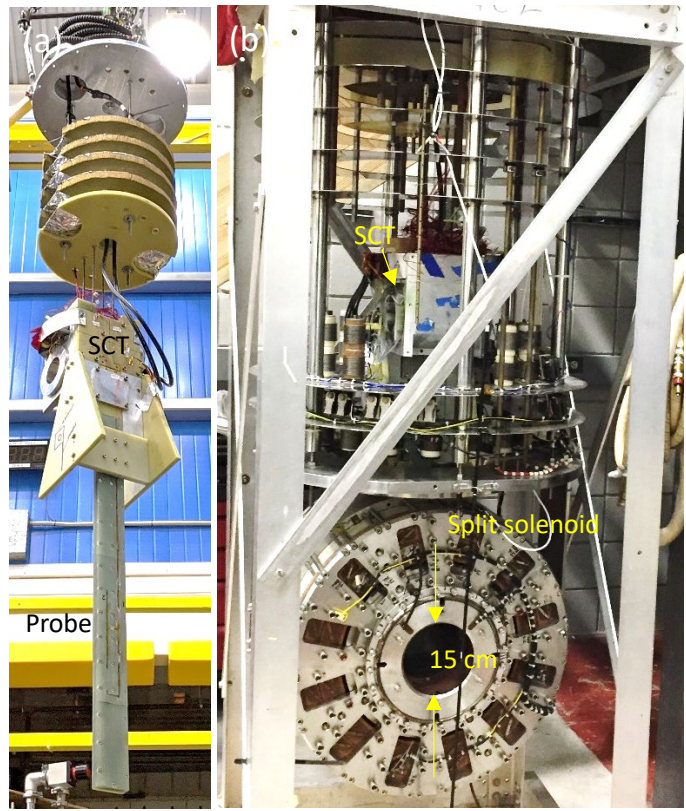

Fig. 3. (a) the probe joined to SCT which is attached to a supporting structure and the top flange to be inserted into the split magnet, and (b) SCT and the probe inside the split magnet assembly with 15 cm bore size.

2.4 The test sequence

The magnetic field is provided by an Oxford 12 T split magnet that has the high field region of 15 cm in diameter and allows maximum probe cross-section of 30 x 70 mm$^2$. The magnet has a homogeneity of 0.2% over 6 cm diameter spherical volume. The applied magnetic field and electrical current polarity is such that the electromagnetic force pushes the two sides of the sample together. A 15 cm long voltage tap was attached to each leg of the sample, utilizing the entire high field region which is greater than the cable's transposition pitch of 11 cm.

In the experiment, both the sample probe and SCT were immersed in liquid helium. The experiment began with ramping the SCT primary current to 20 A (which corresponds to SCT output of about 20 kA) in zero field. This is to check the SCT output current measurement to verify the proper functioning of the system. Subsequently, the magnet was ramped to different fields for $I_c$ testing with a ramp rate of typically 0.025 -0.168 T/min. Before each $I_c$ measurement in field, the SCT was quenched by energizing quench heaters located on the secondary coil of the SCT to eliminate the induced residual current due to the field ramp. Then the primary current was ramped at 0.1 A/s while the output current and voltage in the sample high field region are monitorred until it reaches critical point. In this fashion, the V-I curves were measured for applied field from 2 T to 10 T. The critical currents were determined by a criterion of 0.1 µV/cm. The n values are determined by fitting the E-I curves with the exponential function

$$E = E_c (I/ I_c)^n \qquad (1)$$

where $E_c$ is the criterion.

Furthermore, in order to compare the performance of the cable with one of its strand. A 1.5 meter long strand was extracted from the same LHC cable. The extracted strand was wound on a standard ITER barrel, and its $I_c$ was tested at 4.2 K in a wire test facility designed for ITER strand verification testing [8].

### 3 Results and Discussions
#### 3.1 The electric field vs current (E-I) curves

The total resistance of the two solder joints between the SCT and the sample is 2.0 nΩ obtained by fitting the linear part of the V-I traces taken prior to superconducting-to-normal transition. This joint resistance value is sufficiently low that decay time constant is much longer than the time takes by a critical current measurement of a few minutes. At this level of joint resistance, no significant additional primary current is needed to compensate the the energy dissipated at the joints.

E-I curves up to the superconducting-to-normal transitions in magnetic field were obtained between 2 and 10 T, as shown in Fig. 4. Although many cases quenched before reaching criterion of 0.1 µV/cm, the E-I curves above 4 T show clear superconducting-to-normal transitions. The $I_c$ and $n$ values determined by fitting with equation (1) are plotted in Fig. 5 as a function of applied field.

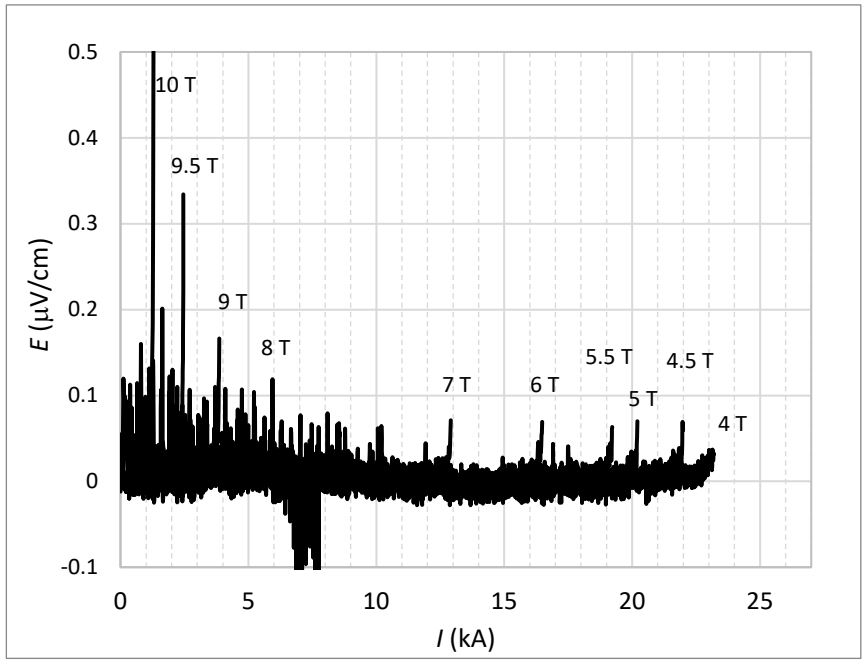

Fig.4. *E - I* curves of the sample obtained in 2 - 10 T magnetic field. A constant background is removed from each curve. The current was measured by one of the two Hall sensors.

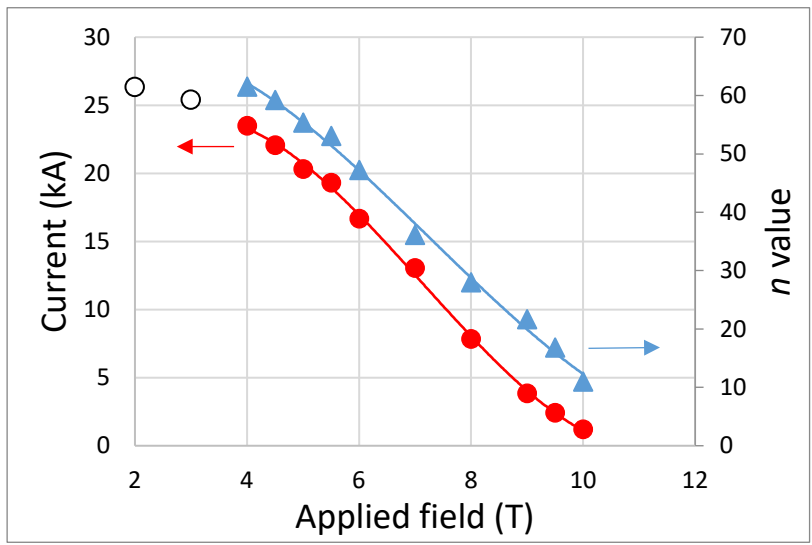

Fig. 5. The measured $I_c$ and *n* value versus applied magnetic field. The solid line are guides to the eye. Two empty points at the low fields side are the current values at which the sample quenched without transitions.

### 3.2 Self-field correction

For a high current density cable at relatively low magnetic fields such as NbTi Rutherford cable, the effect of the self-field is significant. Due to the twist, each strand is subject to a longitudinally changing self-field with a period equal to the cable pitch [9]. Only if the effect of the self-field are quantified, it is possible to determine the cable performances. So this section is concerned with the self-field calculation by finite element method (FEM) using ANSYS software.

The magnet field (B) experienced by a cable can be described as a sum of the applied field $B_a$ and the self-field $B_s$. Fig. 6 plots the calculated $B_s$ distribution in the cross-sectional area of our cable, assuming transport current of 10 kA. As depicted in the figure, the applied field $B_a$ is parallel to the y direction. Since x-component of $B_s$ only make small contribution to the total $B$, only y-component of $B_s$ is ploted in Fig. 6 (b). The self-field in extract strand test was also calculated as shown in Fig. 7.

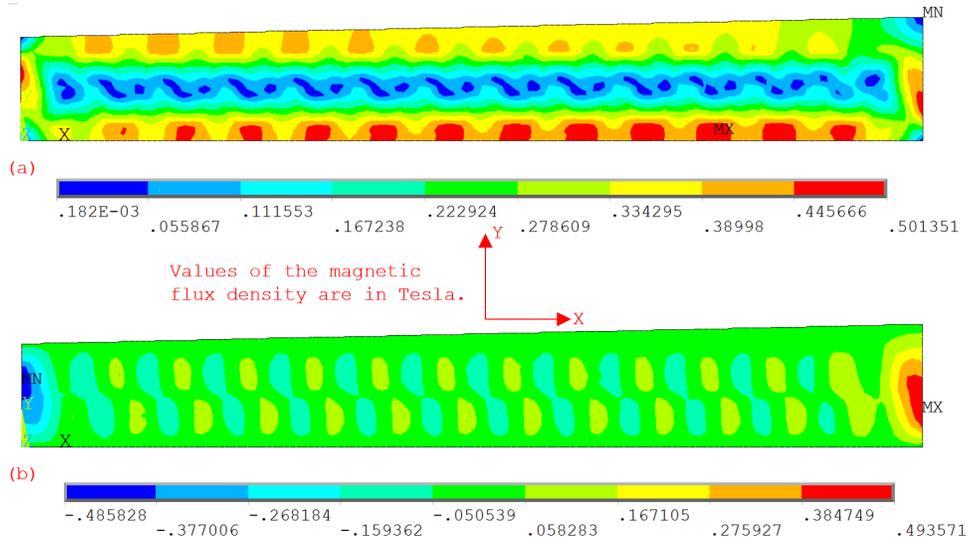

Fig. 6. Self-field $B_s$ distribution in tesla in cable cross-sectional area assuming 10 kA transport current (a) the magnitude (scalar) of the self-field in tesla, and (b) the y-compoment of the self-field in tesla

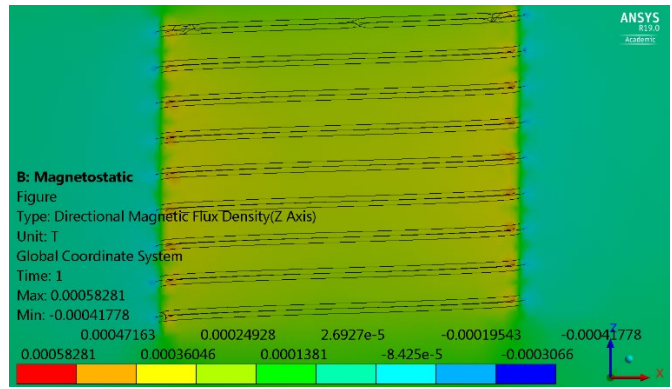

Fig. 7. Calculated self-field of the extracted strand in its critical current test using an ITER barrel assuming 1 A transport current. The component of $B_s$ along the z (axial) direction

Since the self-field is proportional to the transport current, so the self-field at critical current $I_c$ can be expressed as

$$B_s = K*I_c \qquad (2)$$

Where *K* is a constant. An important work on critical current measurements of cables done at BNL concluded that the critical current is only determined by the properties of the superconductor in the peak field region [10]. Our calculation results in $K_{max}$ = 0.583 and 0.501 T/kA respectively for the single strand on ITER barrel and the cable with its broard face perpendicular to the applied field.

For the LHC cable, according to our analysis, the max vector sum of $B_s$ and $B_a$ is almost the same as the sum of the max y component of $B_s$ ($B_{sy}$) and $B_a$, so $K_y$ = 0.4936 T/kA is used for each critical current at each field for self-field correction. For the ITER barrel, in our calculation, $B_a$ is parallel to the z direction, so $K_z$ = 0.583 T/kA is used for each critical current at each field for self-field correction.

Based on above FEM analysis, the applied magnetic fields are corrected by adding the calculated maximum self-field in y direction. After self-field correction, data in Fig. 5 are replotted in Fig. 8. The critical current of the extracted strand, also corrected for self-field, are scaled by multiplying 28, the number of strand in the cable, also presented in Fig. 8 for comparison. It should be noted that at 7 T, the critical current is 14 kA and n value is 36.1, consistent with the values for LHC type 01 cable in the literature [7], [11]. A very good correlation between the cable critical current and the sum of the critical current of extracted strand was observed at CERN [10]. But our difference between the cable $I_c$ and 28 x strand $I_c$ seems to be significant. This may suggest that the extracted strand has better performance than most other strands in the cable.

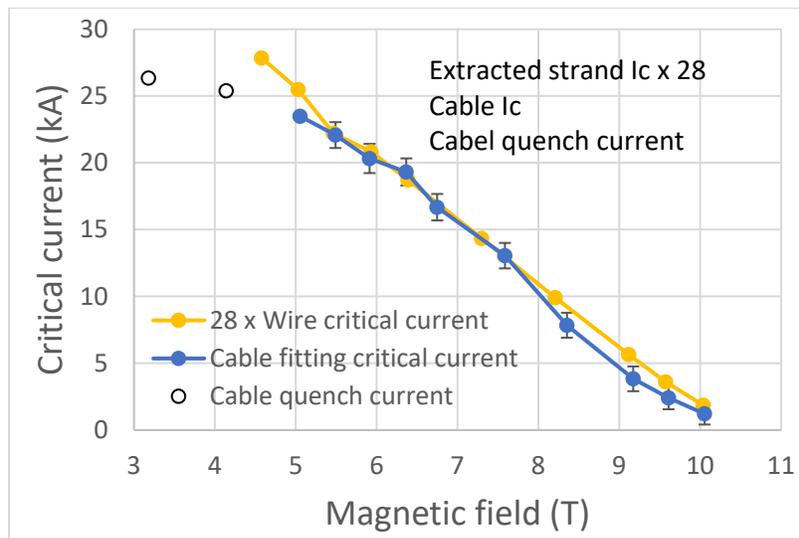

Fig. 8. Comparison of critical currents between the cable and the strand for after self-field correction

### 3.3 Other uncertainties in critical current measurement

Inductive signal from the Rogowski coil is usually integrated to measure output current. However, in this particular experiment, the digital integration of Rogowski coil signal is noisy. Therefore this experiment rely on a carefully calibrated Hall sensor which is robust against electromagnetic noises. In order to further verify the secondary current, the primary current was multiplied by 1000, the transformer ratio. Since the deviation due to energy dissipation at the joints is relatively small, this calculated output current should be very close to the actual output current. Indeed this calculated output based on primary current agrees very well with measured output by the Hall sensor.

A possible source of error is from the stray field from the magnet at the Hall sensors. However, in our experiment configuration, the stray field of the magnet coil is mostly parallel to the sensor which produce negligible Hall signal. This is confirmed by our experiment, that Hall signal is zero regardless the applied magnetic field value when the output current is zero (quenched). The error of measured critical current is from the uncertainties of the Hall sensors calibration, of the distance between the Hall sensors and the output leads, and of the fitting process of V-I curves. Sum of these three errors are displayed as error bar on Fig. 8. The error from Hall sensor calibration is a constant of about 782 A, which results the smaller critical currents measured at high fields would have larger errors. The estimated error at 7 T is about 7%.

### 3.4 Current induced by field ramp

In our test configuration the superconducting cable forms a loop in the high magnetic field region, It is expected that inductive current is significant during a field ramp. As observed, this inductive current is in the order of a few kA, which could introduce a significant error in critical current if the sample current is measured by integration of inductive voltage (Rogowski coil). This is why before each critical current test, the inductive current was quenched by the quench heaters. Instrestingly at high field, this inductive current is comparable with the critical current of our sample, so a superconducting to normal transition could occur due to the inductive current. Fig. 9 is an interesting example of this effect. It displays the induced secondary current by applied field ramp from 9 T to 10.8 T. The induced secondary current causes superconducting-to-normal transitions at roughly 2, 1.2, and 0.8 kA at respective magnetic field of 9.64, 9.9 and 10.1 T, consistent with critical currents at these fields.

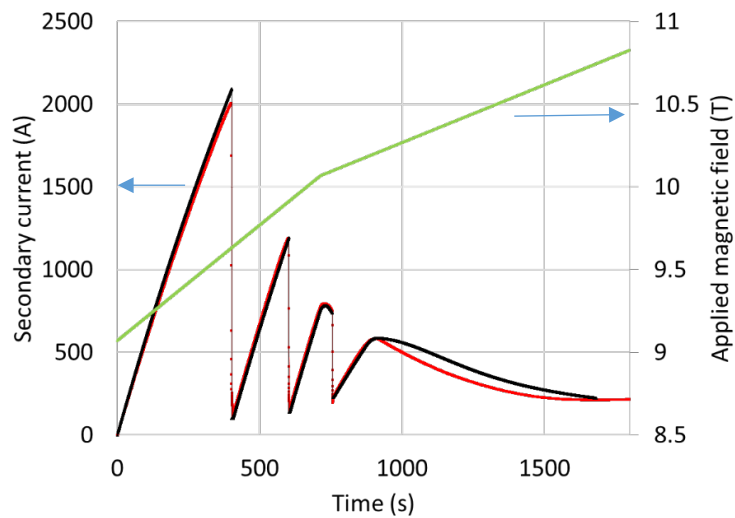

Fig. 9. Induced secondary current by magnetic field ramp from 9 T to 10.8 T. Once the induced current reaches $I_c$, it is self-quenched. The next current ramp starts automatically. As expeccted, the quench current decreases with increasing magnetic field.

## 4  Conclusion

The critical current of a LHC type 01 NbTi Rutherford cable is successfully measured in high magnetic field using the superconducting transformer as high current source. This verifies the correct operation of this 50 kA grade superconducting transformer in high magnetic fields. Both Rogowski inductive coils and Hall sensors are used to measure the SCT output current. FEM modeling was performed to calculate the self-field of these high current measurements. Our measured critical current is in agreement with what in the literature. In addition, the critical current of a strand extracted from the test cable was measured and compared with that of the cable. The reasonable agreement between the two also comfirmed the proper functioning of the SCT.

**Acknowledgement**

We thank Dr. Luca Bottura and Dr. Amalia Ballarino for kindly providing the LHC NbTi cable. We are thankful to Mr. Robert Walsh and Mr. Erick Arroyo for helps on the operation of the split magnet. This work is supported by the user collaboration grant program (UCGP) of the NHMFL which is supported by NSF through NSF-DMR-1157490 and 1644779, and the State of Florida.